# Exotic radiation from a photonic crystal excited by an ultra-relativistic electron beam

N. Horiuchi,[1] T. Ochiai,[2] J. Inoue,[2,3] Y. Segawa,[1] Y. Shibata,[4] K. Ishi,[4] Y. Kondo,[5] M. Kanbe,[5] H. Miyazaki,[5] F. Hinode,[6] S. Yamaguti,[7] and K. Ohtaka[8]

[1]*Photodynamics Research Center, The Institute of Physical and Chemical Research (RIKEN), Sendai 980-0845, Japan*
[2]*Nanomaterials Laboratory, National Institute for Materials Science (NIMS), Tsukuba 305-0044, Japan*
[3]*International Center for Young Scientists, NIMS, Tsukuba 305-0044, Japan*
[4]*Institute of Multidisciplinary Research for Advanced Materials, Tohoku University, Sendai 980-8577, Japan*
[5]*Department of Applied Physics, Tohoku University, Sendai 980-8579, Japan*
[6]*Laboratory of Nuclear Science, Tohoku University, Sendai 982-0826, Japan*
[7]*Graduate School of Science and Technology, Chiba University, Chiba 263-8522, Japan*
[8]*Center for Frontier Science, Chiba University, Chiba 263-8522, Japan*
(Dated: December 20, 2005)

We report the observation of an exotic radiation (unconventional Smith-Purcell radiation) from a one-dimensional photonic crystal. The physical origin of the exotic radiation is direct excitation of the photonic bands by an ultra-relativistic electron beam. The spectrum of the exotic radiation follows photonic bands of a certain parity, in striking contrast to the conventional Smith-Purcell radiation, which shows solely a linear dispersion. Key ingredients for the observation are the facts that the electron beam is in an ultra-relativistic region and that the photonic crystal is finite. The origin of the radiation was identified by comparison of experimental and theoretical results.

PACS numbers: 41.60.-m, 41.75.Ht, 42.70.Qs

Radiations from moving charged particles have attracted much interest in various research fields of physics and materials science. The radiations, which include Synchrotron radiation, Cherenkov radiation, transition radiation, and so on, can serve for a coherent source of radiation, beam diagnostic, particle detection, and a probe of a specimen [1]. Among them, Smith-Purcell radiation (SPR) [2] is a promising candidate to make a compact free-electron laser of arbitrary frequency. In SPR, coherent radiation is emitted from an electron beam scanning just above a diffraction grating. There, the most important thing is that the output frequency is widely tunable by changing the period of the grating [3–11].

SPR also occurs in photonic crystals (PC) [12–16], *i.e.*, multi-dimensional periodic dielectric structures. When we consider a PC as a converter from an electron beam to propagating radiation, various features inherent in the PC will arise by exciting highly confined radiation modes, or in other words, photonic band (PB) modes of PC. Direct excitation of PB modes may result in exotic radiation, which has never been observed in diffraction gratings. One example of such exotic radiation caused by an electron beam interacting with a PC was theoretically predicted by Luo *et al.* [17]. There, the Cherenkov effects reveal a strange directivity depending on the frequency of observed radiation. It should be stressed that PB modes can be controlled by changing various parameters of the PC. Since the radiation is strongly modulated by PB modes, it can be used to probe them.

In this Letter, we report the first experimental observation of exotic radiation from a PC induced by an electron beam. The exotic radiation, which hereafter is called unconventional SPR, is completely different from conventional SPR in the sense that its signal does not appear on the geometrical lines of conventional SPR in frequency-momentum space. Instead, unconventional SPR follows the dispersion relation of PB modes with a certain parity. Since the PB structure extends throughout entire frequency-momentum space, a variety of PB modes can be utilized. To observe unconventional SPR, key ingredients are ultra-relativistic velocity of the electron beam and the broken translational invariance of a finite-sized PC. The radiation is identified by using carefully prepared samples and by excellent agreement of the radiation spectrum with a refined theory of SPR that takes into account the finiteness of the sample. It should be stressed that unconventional SPR is not classified into known radiations involving an electron beam. In particular, the present mechanism of conversion from an evanescent wave to a propagating one is neither by Umklapp scattering in conventional SPR nor by decline in the slope of the light line in Cherenkov radiation.

Let us recapitulate conventional SPR mechanism briefly. We used the coordinate system shown in Fig. 1. Light of frequency $\omega$ emitted by an electron traveling with velocity $v$ along the $x$-axis has a wavevector $(k_x, \pm\Gamma, k_z)$, with the $x$ component given by $k_x = \omega/v$ and the $z$ component $k_z$ being arbitrary. By energy conservation, $\Gamma = \sqrt{(\frac{\omega}{c})^2 - k_x^2 - k_z^2}$. Since $v < c$, $\Gamma$ is pure imaginary, meaning that the emitted light is evanescent and decays with constant $|\Gamma|$ with distance from the trajectory. In other words, the line $\omega = vk_x$, called the $v$ line, lies outside the light line $\omega = ck_x$. When we observe far-field SPR within the $xy$ plane, $k_z$ can be set to zero. A one-dimensional PC having a periodicity $d$ in the $x$ direction then produces SPR by giving a Umklapp change of the integer multiple of $\frac{2\pi}{d}$ to the $x$ component of the wavevector of initial light. Using the same symbol $k_x$ to represent the $x$ component of SPR, we may say that the initial light on the $v$ line is Umklapp-shifted to the new $v$ lines of dispersion $\omega = v(k_x + n\frac{2\pi}{d})$, $n$ being an arbitrary integer. We shall call these lines $v_n$ lines. The light on the $v_n$ line thus has the $y$ component of the wavevector given by $\Gamma_n = \sqrt{(\frac{\omega}{c})^2 - (\frac{\omega}{v} - n\frac{2\pi}{d})^2}$. It reaches a far-field observation point as an SPR signal when this $\Gamma_n$ is real, i.e., when the frequency $\omega$ of the radiation on the $v_n$ line is inside the light cone. This can always happen for a $v_n$ line with positive $n$. In the PCs shown in Fig. 1, PB modes have their band structure in $(k_x, \omega)$ space. When the $v_n$ line crosses a PB dispersion curve, corresponding peaks appear in the SPR spectrum [12, 13, 16]. This is conventional SPR involving a PC.

The experimental setup is depicted in Fig. 1. A short-bunched electron beam of 150 MeV from a linear accelerator at the Laboratory of Nuclear Science (LNS), Tohoku University traveled in the $x$ direction above a PC. The macro and micro pulse widths were 2 $\mu$s and 0.67 ps, respectively. The average beam current was 1.4 $\mu$A. The cross section of the beam was about $10 \times 12$ mm$^2$. The distance between the trajectory and PC surface was kept constant at 10 mm throughout the experiments. The velocity of the electrons was $v = 0.99999c$. With this ultra-relativistic $v$, the decay constant $|\Gamma|$ is very small, being $\omega/c$ times $4.47 \times 10^{-3}$. The millimeter wave from the PCs was collected by mirrors and detected by a helium-cooled Si bolometer. The spectrum of the emitted light was analyzed by a Martin-Puplett-type Fourier-transformation spectrometer. The resolution of the spectrometer was 3.75 GHz.

We used four samples, which are monolayers of arrayed about 30 cylinders of polytetrafluoroethylene (PTFE), fused quartz or aluminum (see Table I). The samples were all PCs with diameter and periodicity in the millimeter range and having periodic in the $x$ direction with $z$-directed cylinder axes. Samples 1, 2 and 3 were dielectric cylinders, whose effective dielectric constants $\epsilon_\text{eff}$ [18] of the monolayer became larger in this order. Sample 4 was a monolayer of metallic cylinders, a PC of perfect conductors in the millimeter wave region.

The radiation signals were collected by sweeping $(\theta, \omega)$ in the $xy$ plane of Fig. 1. By varying $\theta$ in the angle range $60° < \theta < 110°$ and using the relation $k_x = (\omega/c)\cos\theta$, we obtained a radiation intensity map in the $(k_x, \omega)$ space. Figure 2 shows the radiation intensity maps for (a) sample 2 and (b) sample 3. Because $v \simeq c$, the $v$ line overlaps with the light line.

Conventional SPR signals, which show a non-monotonic intensity change by the PB effect, can be clearly seen along the lines $v_1$ and $v_2$ [16]. A series of strong signal lines can also clearly be seen in Fig. 2 off $v_n$ lines, referred to as unconventional SPR. The signals show up along the straight, with slopes definitely smaller than that of the light line. These lines are marked as $u_n$ in both (a) and (b). The slope of $u_1$ in sample 3 is clearly smaller than that in sample 2. In sample 1, for which data are not given here, line $u_1$ further approaches $v_1$ line. Thus, the larger the effective dielectric constant of the monolayer is, the smaller their slopes become. The estimated values of the slopes are shown in Table I. Unconventional SPR was not observed in sample 4 of metal cylinders, as expected (data not shown).

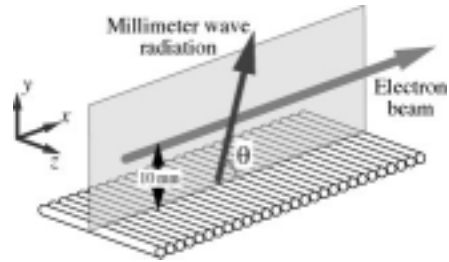

FIG. 1: Experimental geometry. The electron beam is in the $x$ direction 10 mm above the photonic crystal. Cylinders for which data are given in Table I are arrayed with axes in the $z$ direction. The intensity of millimeter-wave radiation is measured as a function of $(\theta, \omega)$ at the far-field observation point in the $xy$ plane.

Taking into account the general tendency between effective dielectric constant and PB frequencies, it is reasonable to assume that unconventional SPR accompanies the excitation of those PBs that are quasi-guided modes with a quantized wavevector component in the thickness direction ($y$ direction in our case) [19]. The conventional theory of SPR, however, predicts only signals on the $v_n$ lines because periodicity is assumed to be infinite. To explain the signals off the $v_n$ lines, therefore, we need to incorporate explicitly the finiteness of the PC length. In order to justify the assumption, we performed improved theoretical calculations in which finite effects of the sample were taken into account [20, 21], in contrast to conventional SPR theory, and compared with the observation.

The results obtained by the improved theory are shown in Figs. 3. Figures 3 show superposed intensity maps of the radiation spectra and PB diagrams of the quasi-guided modes [22]. Figs. 3 (a) and (b) both have unambiguously unconventional signals $u_1 \sim u_5$ that appear in Figs. 2. The slopes and intensity in Figs. 3 reproduced well the experimental results shown in Figs. 2. Strong signals having a Fabry-Perot oscillation are seen along the light line and correspond to light propagation in the $x$ direction. They are strong because the $v$ line is almost the same as the light line when $v \sim c$. Since angle $\theta$ is bounded as $\theta > 60°$, the signals in the forward direction $\theta = 0°$ are out of range in our experimental set up. The agreement between the theory and experiments is quite well, except the strong signal at $(k_x, \omega) = (-0.4 \times 2\pi/d, 130$ GHz$)$ in Fig. 2 (a). It should be noted that the agreement between theoretical calculations and experimental

TABLE I: Characteristics of PC samples

| Sample | Substance of PC | dielectric constant of cylinders | Number of cylinders | Diameter of cylinders | Distance between cylinders | $\epsilon_{\text{eff}}$ | Slope of $u_1$ |
|---|---|---|---|---|---|---|---|
| 1 | PTFE | 2.05 | 14 | 3.1 mm | 6.2 mm | 1.14 | $0.92c$ |
| 2 | PTFE | 2.05 | 28 | 3.1 mm | 3.1 mm | 1.35 | $0.78c$ |
| 3 | Fused quartz | 4.41 | 28 | 3.0 mm | 3.0 mm | 1.79 | $0.59c$ |
| 4 | Aluminum | — | 28 | 3.0 mm | 3.0 mm | — | — |

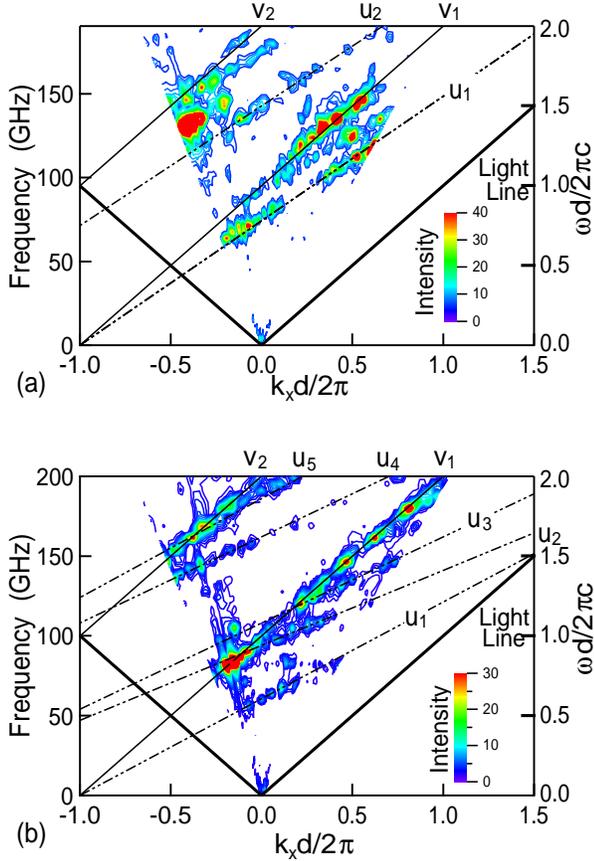

FIG. 2: (Color) Intensity maps of radiation observed in the $xy$ plane for sample 2 (a) and sample 3 (b). Thick solid lines with slopes $\pm c$ are the light lines forming the boundary of the light cone. Two $v_n$ lines, marked as $v_1$ and $v_2$, are drawn parallel to the light line. Dotted lines $u_1 \sim u_5$ indicate the lines on which unconventional SPR appears.

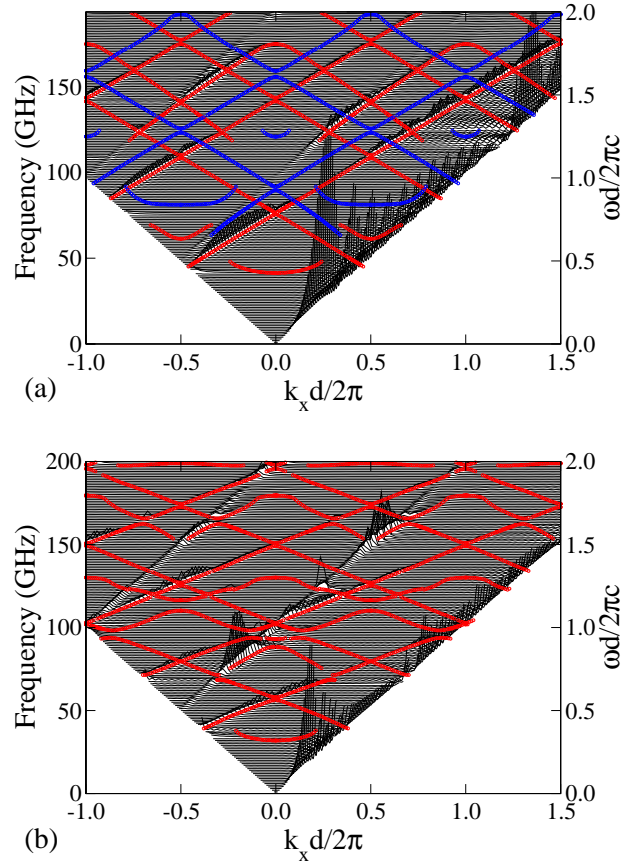

FIG. 3: (Color) Bird's eye views of calculated intensity maps of radiation observed in the $xy$ plane for sample 2 (a) and sample 3 (b). The boundaries of the map are the light lines of slope $\pm c$. The quasi-guided PBs of even (odd) parity with respect to the mirror plane $y = 0$ are also shown by red (blue) circles in (a). In (b) only the even-parity modes are shown. Along the right boundary of the light cone, a strong emission of light can be seen (see text). Unconventional SPR appears precisely along the even-parity dispersion curves, odd-parity PBs being left silent.

results is quite good near the most complicated part of the radiation spectra.

Besides, in Figs. 3, it is remarkable that the unconventional signals appear exactly along the dispersion curves of quasi-guided PBs. Each of the quasi-guided PBs has a definite parity with respect to the mirror plane $y = 0$ of the monolayer and is shown by either red (even parity) or blue (odd parity) circles in Figs. 3. In Fig. 3(b), only the quasi-guided PBs of even parity are shown. As is obvious in Fig. 3(a), only the even parity modes par-ticipate in unconventional SPR. This statement is indeed reconfirmed by Fig. 3(b) of sample 3, which has a PB structure much more complicated than that of sample 2.

In order to further understand unconventional SPR, it is crucial that only mirror-symmetric PB modes produce the signals. As mentioned before, in the ultra-relativistic region, the decay constant $|\Gamma|$ is very small. This implies that the evanescent light of the electron beam behaves as a plane-wave propagating in the $x$ direction, with polarization in the $y$ direction and, most importantly, with amplitude practically constant in the $y$ direction. Thus, the experimental situation is essentially such that a $y$-polarized plane-wave light of $k_x$ enters the PC from the edge. It is well known in light transmission that any PBs of frequency $\omega$ can be excited regardless of the coincidence of their wavevectors with $k_x$. Also, selective excitation of quasi-guided PB modes occurs, depending on the matching of the symmetries between incident light and modes to be excited. In our case, the even-parity PB modes participate because the $y$-independent incident light is mirror-symmetric with respect to the plane $y = 0$.

Gathering up all discussion presented above, we conclude that the observed signals reported in this Letter originate from truly direct excitation of PB. This process, the generation of plane-wave light from an electron beam, is clearly different from well-known ones such as Cherenkov radiation and SPR. In the former case, a propagating light is generated owing to the decline in the slope of the light line, and in the latter case, Umklapp scattering is responsible for the light generation.

It should be noted that our PC samples have lattice constants of several millimeters, and thus resultant radiation is in the millimeter wave region. One of the key ingredients of unconventional SPR is the usage of dielectric substances instead of metallic ones. A metal in this region can be regarded as a perfect conductor, which does not support the surface plasmon on its boundary. This is the main reason why unconventional SPR is absent in sample 4. So far, several groups have reported millimeter-wave SPR from metallic diffraction gratings [5, 6, 9]. They were all conventional signals. However, if the gratings are downsized so that SPR of the visible or ultra-violet range is generated, we would observe unconventional SPR caused by excitation of the PBs of the surface plasmon.

In conclusion, exotic radiation (unconventional SPR) due to direct excitation of PBs was observed from finite-sized PCs. The evanescent light induced by an electron beam with ultra-relativistic velocity enters from the edge of the dielectric PC and is converted into plane-wave light. A selective mode excitation occurs depending on the matching of symmetries between incident evanescent light and PBs. The process by which unconventional SPR is generated is different from that of Cherenkov radiation or conventional SPR. The spectrum of unconventional SPR is widely tunable by changing various parameters of the PC. In addition to the application of a coherent light source, unconventional SPR also opens up a possibility of probing the PB structure with high accuracy.

The authors thank Messrs. Tsutomu Tsutaya, Yuki Chiba and Ryosuke Watanabe and the staff of LNS, Tohoku University for their help in the experiments. This work was supported by the Special Coordination Funds for Promoting Science and Technology from the Ministry of Education, Culture, Sports, Science, and Technology of Japan.

[1] M. L. Ter-Mikaelian, *High-Energy Electromagnetic Processes in Condensed Media* (Wiley-Interscience, New York, 1972).
[2] S. J. Smith and E. M. Purcell, Phys. Rev. **92**, 1069 (1953).
[3] G. Toraldo di Francia, Nuovo Cimento **16**, 61 (1960).
[4] P. M. van den Berg, J. Opt. Soc. Am. **63**, 1588 (1973).
[5] A. Gover, P. Dvorkis, and U. Elisha, J. Opt. Soc. Am. B **1**, 723 (1984).
[6] G. Doucas, J. H. Mulvey, M. Omori, J. E. Walsh, and M. F. Kimmitt, Phys. Rev. Lett. **69**, 1761 (1992).
[7] O. Haeberlé, P. Rullhusen, J.-M. Salomé, and N. Maene, Phys. Rev. E **49**, 3340 (1994).
[8] J. E. Walsh, J. K. Woods, and S. Yeager, Nucl. Instrum. Methods, Phys. Res. Sec. A **341**, 277 (1994)
[9] J. K. Woods, J. E. Walsh, R. E. Stoner, H. G. Kirk, and R. C. Fernow, Phys. Rev. Lett. **74**, 3808 (1995).
[10] J. H. Brownell, J. E. Walsh, and G. Doucas, Phys. Rev. E **57**, 1075 (1998).
[11] S. R. Trotz, J. H. Brownell, J. E. Walsh, and G. Doucas, Phys. Rev. E **61**, 7057 (2000).
[12] K. Ohtaka and S. Yamaguti, Opt. Spectrosc. **91**, 477 (2001).
[13] S. Yamaguti, J. Inoue, O. Haeberlé, and K. Ohtaka, Phys. Rev. B **66**, 95202 (2002).
[14] F. J. García de Abajo A. G. Pattantyus-Abraham, N. Zabala, A. Rivacoba, M. O. Wolf, and P. M. Echenique, Phys. Rev. Lett. **91**, 143902 (2003).
[15] F.J. García de Abajo and L.A. Blanco, Phys. Rev. B **67**, 125108 (2003).
[16] K. Yamamoto, R. Sakakibara, S. Yano, Y. Segawa, Y. Shibata, K. Ishi, T. Ohsaka, T. Hara, Y. Kondo, H. Miyazaki, F. Hinode, T. Matsuyama, S. Yamaguti, and K. Ohtaka, Phys. Rev. E **69**, 045601 (2004).
[17] C. Luo, M. Ibanescu, S. G. Johnson, and J. D. Joannopoulos, Science **299**, 368 (2003).
[18] J. C. Maxwell-Garnett, Philos. Trans. R. Soc. London, Ser. A **203**, 385 (1904).
[19] K. Ohtaka in *Photonic Crystals* edited by K. Inoue, and K. Ohtaka, (Springer, Berlin, 2004), p. 54 and p. 84.
[20] T. Ochiai and K. Ohtaka, Phys. Rev. B **69**, 125106 (2004).
[21] T. Ochiai and K. Ohtaka, submitted to Phys. Rev. E.
[22] K. Ohtaka, J. Inoue, and S. Yamaguti, Phys. Rev. B **70**, 035109 (2004).